\documentclass[11pt]{article}
\usepackage{cite}
\usepackage{amsmath,amsfonts,amssymb}
\usepackage[small,bf,hang]{caption}
\usepackage{graphicx}

\def\hybrid{
        \topmargin -20pt
        \oddsidemargin 0pt
        \headheight 0pt \headsep 0pt
        \textwidth 6.25in 
        \textheight 9.5in 
        \marginparwidth .875in
        \parskip 5pt plus 1pt \jot = 1.5ex}

\hybrid

 \csname
@addtoreset\endcsname{equation}{section}

\def\moth{\mathsurround=0pt}
\newdimen\zo \zo=0pt

\def\tick{\leaders\hrule height 0.5ex depth 0pt \hskip 0.5pt}
\def\upboxfill{$\moth \setbox\zo\hbox{\tick}%
  \hskip 3pt\hbox to 0pt{$\tick$\hss}\hrulefill \hbox to 7.5pt{$\tick$\hss}$}

\def\dtick{\leaders\hrule height .34pt depth 0.5ex \hskip 0.5pt}
\def\downboxfill{$\moth \setbox\zo\hbox{\dtick}%
  \hskip 2pt\hbox to 0pt{$\dtick$\hss}\hrulefill \hbox to 2pt{$\dtick$\hss}$}

\def\bec{\begin{center}}
\def\ec{\end{center}}

\def\be{\begin{equation}}
\def\ee{\end{equation}}
\def\bea{\begin{eqnarray}}
\def\eea{\end{eqnarray}}
\def\ba{\begin{array}}
\def\ea{\end{array}}

\begin{document}

\begin{titlepage}
\begin{center}
\vskip 2.5cm
{\Large \bf {
Magnons and spikes for $\mathcal{N}=2$ linear quivers and their non-Abelian T-duals}}\\
\vskip 1.0cm
{\large {Dibakar Roychowdhury}}
\vskip 1cm
{\it {Department of Physics}}\\
{\it {Indian Institute of Technology Roorkee}}\\
{\it {Roorkee 247667, Uttarakhand, India}}\\

\vskip 2.5cm
{\bf Abstract}
\end{center}

\vskip 0.1cm

\noindent
\begin{narrower}
We compute the spectra associated with various semiclassical string states that propagate over $\mathcal{N}=2$ Gaiotto-Maldacena backgrounds. As an interesting special case, for the Abelian T- dual solution, we discover giant magnon and single spike configurations while imposing appropriate boundary conditions. However, for Sfetsos-Thompson backgrounds, one has to adopt a different string embedding which reveals ``modified'' dispersion relations both for magnons and spikes. These results boil down into the standard dispersion relations in the limit when the rank of the associated $SU(N_c)$ color gauge group becomes large enough. We further generalize our analysis in the presence of flavor D6 branes. Our analysis reveals a new set of dispersion relations, which shows that both magnons and spikes are ceased to exist in the presence of flavor excitations.
\end{narrower}

\end{titlepage}

\newpage

\tableofcontents
\baselineskip=16pt

\section{Introduction and Overview}
A systematic understanding of type IIB superstring dynamics on $ AdS_5 \times S^5 $ \cite{Berenstein:2002jq}-\cite{Gubser:2002tv} and the associated quantum corrections \cite{Frolov:2002av} has been a challenge for decades. A major input to this problem comes through the discovery of the underlying integrable structure on both sides of the duality \cite{Bena:2003wd}-\cite{Minahan:2002ve}. This opens up a new era in gauge/string duality, including several finer details to the spectrum namely the discovery of mgnon \cite{Hofman:2006xt}-\cite{Dorey:2006dq} as well as spiky excitations\footnote{See, \cite{Arutyunov:2006gs}-\cite{Park:2010vs} for a related literature.}\cite{Kruczenski:2004wg}.

Parallel to these developments, there has been a new development in our understanding of gauge/string duality based on the notion of the non-Abelian T-duality (NATD) \cite{delaOssa:1992vci}-\cite{Maldacena:2000mw} which gives rise to a new class of type IIA geometries starting from its parent $ AdS_5 \times S^5 $ background. Each of these classes of $ \mathcal{N}=2 $ geometries has their own holographic interpretations, which naturally opens up a new realm in the celebrated gauge/string correspondence.

Typically, these geometries could be classified into the following two categories. In the first category, type IIA backgrounds are obtained following a NATD along $ SU(2) $ of the internal $ S^5 $ \cite{ReidEdwards:2010qs}-\cite{Lozano:2016kum}, which gives rise to a class of $ \mathcal{N}=2 $ supergravity backgrounds, known as the Gaiotto-Maldacena (GM) class of geometries \cite{Gaiotto:2009gz}, that are dual to $ \mathcal{N}=2 $ linear quivers living in four dimensions \cite{Gaiotto:2009we}. On the other hand, in the second category, one generates a class of type IIA backgrounds, following a NATD inside $ S^2 $ of $ AdS_5 $. These geometries are conjectured to be dual to $ \mathcal{N}=2 $ matrix models in ($ 0+1 $)d, in the presence of a (massive) irrelevant deformation \cite{Lozano:2017ole}-\cite{Roychowdhury:2023hvq}. Both of these theories are less explored compared to their parent $ AdS_5 \times S^5 $ cousins\footnote{See, \cite{Itsios:2017nou}-\cite{Lozano:2022vsv} for a related literature.}. However, it is the first category of theories that we focus in the present paper.

GM class of geometries include the Abelian T-dual (ATD) as well as NATD (also known as Sfetsos-Thompson (ST) backgrounds \cite{Sfetsos:2010uq}) of $ AdS_5 \times S^5 $ as a part of its landscape. These geometries are dual to $ \mathcal{N}=2 $ linear quivers of ``infinite'' size, whose QFT interpretation is not clear at the moment. These quivers are incomplete in the sense they carry an infinite number of color nodes. The ``completion'' of these $ \mathcal{N}=2 $ SCFTs is achieved following an addition of flavor D6 branes in the bulk \cite{Lozano:2016kum}. A recent investigation reveals that GM class of geometries are integrable (in the Liouvillian sense \cite{Nunez:2018qcj}) only for a sub-sector that includes the ATD and ST background. On the other hand, these backgrounds (and hence the dual $ \mathcal{N}=2 $ SCFTs) fail to be integrable in the presence of flavor degrees of freedom. This naturally raises questions about other finer details, like the existence of magnon and spiky excitations of $ \mathcal{N}=2 $ SCFTs in four dimensions which is similar in spirit to that of $ \mathcal{N}=4 $ SYM in four dimensions.

The purpose of the present paper is to address the question whether there exists magnons \cite{Hofman:2006xt}-\cite{Dorey:2006dq} or spikes \cite{Kruczenski:2004wg} for $ \mathcal{N}=2 $ SCFTs in four dimensions and the answer turns out to be positive under certain circumstances. We show that magnon/spiky dispersion relation(s) corresponds to choosing different embeddings for tpye IIA strings in the bulk geometry. For ATD, these semi classical strings wrap the holographic $ \eta $- direction. On the other hand, for ST they show no dynamics along the holographic direction. On top of it, for ST solution, one finds a class of magnon/ spiky configurations that are characterised by the rank of the color group $ SU(PN_c) $, where $P=1,2, \cdots$ characterizes the rank of different color nodes of $\mathcal{N}=2$ quiver \cite{Lozano:2016kum}. 

For generic ST background, we obtain a class of ``deformed'' giant magnon dispersion relations. These are boiled down into the usual giant magnon dispersion relation \cite{Hofman:2006xt} in the limit when the rank of the color $ SU(PN_c) $ gauge group of the quiver goes to infinity. The same conclusion holds for spikes as well. To summarize, as the rank of the color nodes increases, these additional effects that we denote as ``impurity'' contributions, are naturally suppressed 
\begin{align}
E-J= \frac{\sqrt{\lambda}}{\pi} \sin \frac{p}{2}\Big(1+\mathcal{O}(1/(PN_c)^2)\Big)
\end{align}
where the ``giant magnon'' spectrum is obtained in the large rank ($P N_c  \rightarrow \infty $) limit.
Here, $P N_c \sim \eta_c$ stands for the rank of the associated color nodes in the $\mathcal{N}=2$ linear quiver. In this paper, we estimate corrections to the giant magnon spectrum up to LO in $1/(P^2N^2_c)$ expansion. 

When flavor branes are added, we discover yet another class of dispersion relation which reveals a richer structure. The spectrum could be schematically expressed as
\begin{align}
E-J \sim \frac{\sqrt{\lambda}}{\pi}f(P) \sin \frac{p}{2}\Big(1-\frac{1-f(P)}{f(P)}h(\epsilon,p)  \Big)
\end{align}
where $\eta=\eta_c = P$ stands for the location of the flavor D6 branes and $h(\epsilon ,p)$ is a function of magnon momentum ($p$). As our analysis reveals, the function $f(P)\sim \frac{1}{(P- \eta_c)^{3/2}}$ appears to be singular as the string approaches the flavor branes. This clearly indicates that the magnon states are not well defined in the vicinity of flavor nodes in $\mathcal{N}=2$ quiver. On a similar note, one finds that the spiky excitations are also ceased to exist in the presence of flavor nodes.

If we assemble all the above information together, then we are tempted to conjecture the following- (i) In the case of ATD example, the associated $\mathcal{N}=2$ quiver acts like a perfect integrable spin chain whose excitations are giant magnons, (ii) For ST/NATD model, the dual $\mathcal{N}=2$ quiver could be thought of as a collection of infinite number of integrable spin chains, where each color node could be thought of equivalent to a spin chain which has excitations similar to that of a giant magnons with some deformations. However, these deformations go away as the rank of the color node becomes large enough. Finally, in the presence of flavor branes we loose all these nice description and the dual quiver ceases to be integrable. The above picture fits perfectly with our previous observation that the ATD and the ST backgrounds are Liouvillian integrable \cite{Nunez:2018qcj}, while the backgrounds with flavor D6 branes are not. Besides, we now have a deeper understanding of the fact that the above integrablity could be modeled in terms of $\mathcal{N}=2$ ``integrable'' spin chains, quite similar in spirit to that of $\mathcal{N}=4$ SYM.

The organisation for the rest of the paper is as follows. In Section 2, we provide all necessary details about the GM class of geometries and the associated sigma model. In Section 3, we discuss rotating string solutions and in particular discuss the magnon and spiky limits of it for ATD and ST backgrounds. In Section 4, we discuss the fate of magnon and spiky excitations in the presence of flavor degrees freedom. Finally, we draw our conclusion in Section 5.
\section{Sigma model and the $\mathcal{N}=2$ background}
We start by considering (semi)classical stringy dynamics over generic Gaiotto-Maldacena (GM) backgrounds. These type IIA backgrounds preserve $ \mathcal{N}=2 $ SUSY which are therefore dual to $ \mathcal{N}=2 $ superconformal quivers in 4D \cite{Gaiotto:2009we}-\cite{Gaiotto:2009gz}. These solutions are typically expressed by means of a potential function\footnote{For the purpose of our present analysis, we focus only on the NS-NS sector of the full type IIA solution.} $ V(\sigma, \eta) $ \cite{ReidEdwards:2010qs}-\cite{Nunez:2018qcj},
\begin{align}
\label{e1}
&ds^2_{IIA}= 4f_1 (\sigma , \eta)ds^2_{AdS_5} +f_2 (\eta , \sigma)(d\sigma^2 +d\eta^2)+f_3(\eta , \sigma)d\Omega_2 (\chi , \xi)+f_4 (\sigma , \eta)d\beta^2\\
&B_2 =f_5 (\sigma , \eta)\sin\chi d\chi \wedge d\xi\\
&ds^2_{AdS_5}=-dt^2 \cosh^2\rho + d\rho^2 + \sinh^2\rho (d\theta^2 +\cos^2\theta d\phi^2_1 +\sin^2\theta d\phi^2_2)\\
&f_1 =\left(\frac{2\dot{V}-\ddot{V}}{\partial^2_{\eta}V} \right)^{1/2}~,~ f_2 = f_1\frac{2\partial^2_{\eta}V}{\dot{V}}~,~f_3 = f_1\frac{2\partial^2_{\eta}V \dot{V}}{\Delta}\\
&f_4 =f_1 \frac{4\partial^2_{\eta}V \sigma^2}{2\dot{V}-\ddot{V}}~,~f_5 = 2\left(\frac{\dot{V}\partial_{\eta}\dot{V}}{\Delta}-\eta \right),~\Delta = (2\dot{V}-\ddot{V})\partial^2_{\eta}V +(\partial_{\eta}\dot{V})^2.
\end{align}

The potential function $V(\sigma , \eta)$ satisfies the Laplace's equation\footnote{Here we denote the derivatives as, $ \dot{V}=\sigma\partial_{\sigma}V $ and $ \ddot{V}=\sigma^2 \partial^2_{\sigma}V +\sigma \partial_{\sigma}V $.} 
\begin{eqnarray}
\partial_{\sigma}(\sigma \partial_{\sigma}V)+\sigma \partial^2_{\eta}V=0~;~R (\eta)=\sigma \partial_{\sigma}V(\sigma, \eta)|_{\sigma =0}
\label{e6}
\end{eqnarray}
where $ R (\eta) $ denotes the charge density and/or the rank of the gauge group.

The correct quantization for charges, enforces certain constraint on the charge density $ R (\eta) $ namely it must vanish at $ \eta =0 $ and should be a piece wise linear function of the form $ R(\eta) = a_i \eta +q_i $, with slope $ a_i $ \cite{ReidEdwards:2010qs}-\cite{Aharony:2012tz}. The slope changes in accordance to the change in the position of the flavor $ D6 $ branes that enforces a \emph{completion} of the bulk geometry. 

As we discuss below, the cases with ATD and the NATD (also known as the Sfetsos-Thompson (ST) solution) respectively correspond to zero and constant slope for the rank function $R(\eta)$ (see Fig.\ref{atd} and Fig.\ref{natd}). These rank functions characterize $\mathcal{N}=2$ quivers with an infinite size and the associated charge density grows to infinity. On the other hand, the presence of flavor D6 branes truncates these infinitely large quivers to a finite length and therefore corresponds to a finite charge density for the boundary theory.  For such configurations, one naturally ensures the boundary conditions, $R(0)=0= R (N_c) $ \cite{Lozano:2016kum}.

Given the background \eqref{e1}, we note down the bosonic sigma model Lagrangian as
\begin{align}
S_P = -\frac{\sqrt{\lambda}}{4 \pi}\int d^2 \sigma \mathcal{L}_P.
\end{align}

Here, we denote the world-sheet coordinates as, $\sigma^{\alpha}=(\sigma^0 , \sigma^1)$ and $\lambda (\gg 1)$ stands for the t'Hooft coupling. The Polyakov Lagrangian could be formally expressed as
\begin{align}
\label{e2.8}
\mathcal{L}_P = \eta^{\alpha \beta}G_{MN}\partial_\alpha X^M \partial_\beta X^N -\varepsilon^{\alpha \beta}B_{MN}\partial_\alpha X^M \partial_\beta X^N
\end{align}
where $\eta^{\alpha \beta}=(-1 , 1)$ denotes the world-sheet metric in the conformal gauge and $\varepsilon^{01}=1=-\varepsilon^{10}$ is the 2D Levi-Civita on the world-sheet.

In what follows, we restrict the stringy dynamics only to the internal space which is subjected to the realisation, $\rho =0$. In other words, we focus on the following subspace
\begin{align}
\label{e2.9}
&ds^2= -4f_1 (\sigma , \eta)dt^2 +f_2 (\eta , \sigma)(d\sigma^2 +d\eta^2)+f_3(\eta , \sigma)d\Omega_2 (\chi , \xi)+f_4 (\sigma , \eta)d\beta^2\\
&B_2 =f_5 (\sigma , \eta)\sin\chi d\chi \wedge d\xi.
\label{e2.10}
\end{align}

Using \eqref{e2.9}-\eqref{e2.10}, one could further simplify \eqref{e2.8} as
\begin{align}
&\mathcal{L}_P =4f_1 (\dot{t}^2 -t'^2)-f_2 (\dot{\sigma}^2 -\sigma'^2)-f_2 (\dot{\eta}^2 -\eta'^2)-f_3 (\dot{\chi}^2 -\chi'^2)\nonumber\\
&-f_3 \sin^2\chi (\dot{\xi}^2 -\xi'^2)-f_4 (\dot{\beta}^2 -\beta'^2)-2f_5 \sin\chi (\dot{\chi}\xi' -\dot{\xi}\chi')
\end{align}
where dot corresponds to derivative with respect to world-sheet time ($ \sigma^0 $) and prime corresponds to derivative with respect to spatial periodic coordinate $ \sigma^1 $.
\section{Rotating strings on $ S^2 $}
In the first place, we look for single spin rigidly rotating string solutions, where the string solition is considered to be rotating on $ S^2 $ of the internal manifold. We choose an ansatz
\begin{align}
\label{e3.1}
t = \kappa \sigma^0 ~,~ \eta = \eta (\zeta)~,~\chi = \chi (\zeta)~,~\xi (\zeta)= \omega \sigma^0 +g(\zeta)~,~\beta =\beta (\zeta)~,~\sigma =\sigma (\zeta)
\end{align}
where we define the function, $ \zeta= a \sigma^0 + b \sigma^1 $.

Clearly, the string is stretched along the holographic axis ($ \eta $) and is simultaneously spinning along the isometry of $ S^2 $ with an angular frequency $ \omega $. Using \eqref{e3.1}, one finds
\begin{align}
\label{e3.2}
&\mathcal{L}_P = 4 \kappa^2 f_1 - f_2 (a^2-b^2) (\sigma'^2 (\zeta)+\eta'^2(\zeta)) -f_3 (a^2 -b^2) \chi'^2(\zeta)-f_4 (a^2 -b^2) \beta'^2(\zeta) \nonumber\\
&- f_3 \sin^2\chi ((\omega + a g'(\zeta))^2 -b^2 g'^2(\zeta)) + 2 \omega b  f_5 \sin\chi \chi'(\zeta)
\end{align}
where prime denotes derivatives with respect to the arguments of the respective functions.
\subsection{Abelian T-dual background}
We start by considering the simplest case of Abelian T-dual (ATD) of $ AdS_5 \times S^5 $. The corresponding potential function reads as \cite{Lozano:2016kum}
\begin{eqnarray}
\label{e3.3}
V_{ATD}(\sigma , \eta)=\log\sigma -\frac{\sigma^2}{2}+\eta^2 ~;~R (\eta)=1
\end{eqnarray}
where the background is created due to $ N_c $ color D4 branes stretched between NS5 branes.

\begin{figure}
\begin{center}
\includegraphics[scale=0.73]{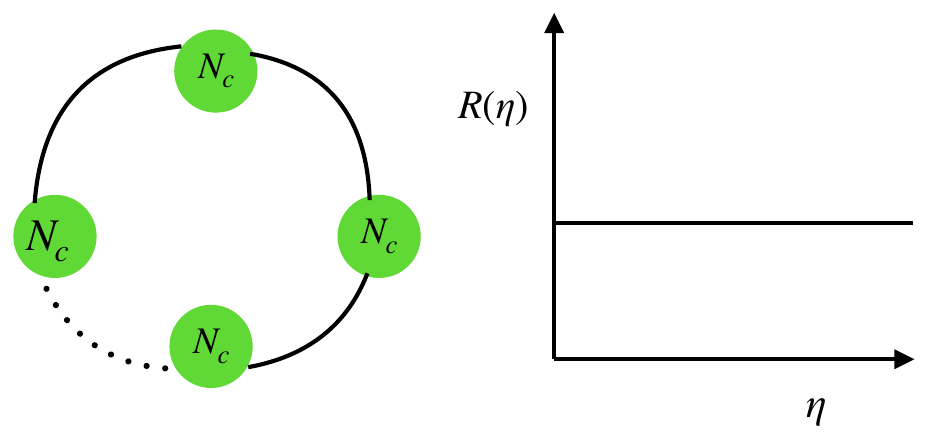}
\end{center}
  \caption{$ \mathcal{N}=2 $ quiver dual to ATD background and the corresponding rank function.} \label{atd}
\end{figure}

Clearly, the above background \eqref{e3.3} represents an infinitely extended flat quiver with constant rank for the associated color group \cite{Lozano:2016kum}. However, these quivers do not meet the criteria $ R(0)=0 =R(N_c)$ (see Fig.\ref{atd}). The $ \eta $ direction, being periodic for the ATD case \cite{Lozano:2016kum}, acts like a compact isometry. This is reflected in the metric functions\footnote{One could check that the above statement is true for any generic values of $ \sigma $.} $ f_i(i=1,2,3,4) $ 
\begin{align}
&f_1(\sigma \sim 0, \eta)=1~,~f_2(\sigma \sim 0, \eta)=4~,~f_3(\sigma \sim 0, \eta)=1\nonumber\\
&f_4(\sigma \sim 0, \eta)=0~,~f_5(\sigma \sim 0, \eta)=-2 \eta.
\end{align}

To proceed further, we adopt a special embedding of the form
\begin{align}
\label{eta}
\eta (\zeta)=\zeta = a \sigma^0 + b \sigma^1~,~\sigma (\zeta)=0.
\end{align}

With this choice \eqref{eta}, the corresponding Lagrangian density could be recast as
\begin{align}
&\mathcal{L}_P=4 \kappa^2 -4 (a^2-b^2) -(a^2 -b^2)\chi'^2(\zeta) \nonumber\\
&-\sin^2\chi ((\omega + a g'(\zeta))^2 -b^2 g'^2(\zeta))-4\omega b \cos\chi.
\end{align}

Next, we note down the equation of motion corresponding to $ g(\zeta) $ 
\begin{align}
g'(\zeta)=\frac{1}{(b^2 -a^2)}\Big( \omega a -\frac{C}{\sin^2 \chi} \Big)
\label{e3.7}
\end{align}
where $ C $ is a constant of integration.

One should therefore think of the $ \eta $- direction as compact isometry of the ATD target spacetime \cite{Lozano:2016kum}, such that the string wraps it and the associated winding number is $ b $. With this choice, the equation for $ \chi(\zeta) $ follows from Virasoro constraints
\begin{align}
\label{e3.8}
T_{\sigma^0 \sigma^0}+T_{\sigma^1 \sigma^1}+2T_{\sigma^0 \sigma^1}=0
\end{align}
which by virtue of \eqref{e3.7} and \eqref{eta} yields
\begin{align}
\label{e3.9}
\chi'^2 (\zeta)=\frac{\omega^2 b^2}{(b^2 - a^2)^2 \sin^2 \chi}\Big( -\sin^4\chi +\frac{D^2}{\omega^2 b^2}\sin^2\chi -\frac{C^2}{\omega^2 b^2}  \Big)
\end{align}
where $ D^2 = 4(\kappa^2-(b+a)^2) (b-a)^2 +2\omega C b $ is a new constant of the theory. 

As a next step, we factorise \eqref{e3.9} as
\begin{align}
\label{e3.10}
\chi'^2 (\zeta)=\frac{\omega^2 b^2}{(b^2 -a^2)^2}\frac{1}{\sin^2 \chi}(\sin^2 \chi_{max}-\sin^2\chi)(\sin^2\chi -\sin^2\chi_{min})
\end{align}
where the following identities hold true
\begin{align}
& \sin^2 \chi_{max}+\sin^2\chi_{min}=\frac{D^2}{\omega^2 b^2}\\
& \sin^2 \chi_{max}\sin^2\chi_{min}=\frac{C^2}{\omega^2 b^2 }.
\end{align}

Here, $ \chi_{max} $ and $ \chi_{min} $ correspond to extremal values of $ \chi $ such that $ \chi'=0 $. The size of the magnon or the spike is determined by the choice of $ \chi_{max} $. The infinite size limit corresponds to setting, $ \sin\chi_{max}=1 $ in the bulk \cite{Arutyunov:2006gs}-\cite{Park:2010vs}. In the dual $ \mathcal{N}=2 $ quivers, this would correspond to magnons with large angular momenta with finite angular difference/linear momentum ($ p $) and spikes with large angular difference and finite angular momentum.
\paragraph{Giant magnons.} We first search for magnon like excitations for the ATD model considered above. We consider heavy magnon states in the dual $ \mathcal{N}=2 $ SCFTs carrying an infinite angular momentum ($ J $). This amounts of setting
\begin{align}
\label{e3.13}
\partial_{\sigma^1}\xi\Big|_{\chi = \chi_{max}}  =0
\end{align}
in the bulk \cite{Park:2010vs} which yields the maximum angle associated with the string soliton as
\begin{align}
\label{e3.15}
\sin^2\chi_{max}=\frac{C}{\omega a}
\end{align}
where the giant magnons correspond to setting, $ C=\omega a $.

Using \eqref{e3.15}, one finds the minimum angle 
\begin{align}
\label{e3.16}
\sin^2\chi_{min}= \frac{a^2}{b^2 }.
\end{align}

Below we note down the energy and the angular momentum of the string soliton
\begin{align}
\label{E}
&E=\frac{\sqrt{\lambda}}{\pi}\int_{\chi_{min}}^{\pi/2}\frac{\cos^2\chi_{min}\sin\chi}{\cos\chi\sqrt{\sin^2\chi - \sin^2\chi_{min}}}\\
&J=\frac{\sqrt{\lambda}}{\pi}\int_{\chi_{min}}^{\pi/2}\frac{d\chi}{\chi'}\sin^2\chi \dot{\xi}=\frac{\sqrt{\lambda}}{\pi}\int_{\chi_{min}}^{\pi/2}\frac{\sin\chi (\sin^2\chi - \sin^2\chi_{min})}{\cos\chi \sqrt{\sin^2\chi - \sin^2\chi_{min}}}
\label{J}
\end{align}
where we use \eqref{e3.16} and scale the above entities by an overall pre-factor $ b $.

Using \eqref{E}-\eqref{J}, the dispersion relation finally turns out to be
\begin{align}
\label{e3.17}
E-J = \frac{\sqrt{\lambda}}{\pi}\int_{\chi_{min}}^{\pi/2}\frac{\sin\chi \cos\chi}{\sqrt{\sin^2\chi - \sin^2\chi_{min}}}d\chi =  \frac{\sqrt{\lambda}}{\pi}\cos \chi_{min}.
\end{align}
where we set, $ \kappa = \frac{1}{4} $ and $ \omega =1 $ without any loss of generality.

The angular difference between the end points of the string soliton turns out to be
\begin{align}
\label{e3.18}
\Delta \xi =2 \int_{\chi_{min}}^{\pi/2}\frac{\sin \chi_{min}\cos\chi}{\sin\chi \sqrt{\sin^2\chi - \sin^2\chi_{min}}}d\chi =  2 \sin^{-1}(\cos \chi_{min}).
\end{align}

Combining \eqref{e3.17} and \eqref{e3.18}, we find the single spin giant magnon dispersion relation
\begin{align}
\label{e3.20}
E-J =\frac{\sqrt{\lambda}}{\pi} \sin \frac{p}{2}
\end{align}
where the geometric angle ($\Delta \xi $) is identified with the magnon momentum ($ p $)  \cite{Hofman:2006xt}. 

\begin{figure}
\begin{center}
\includegraphics[scale=0.73]{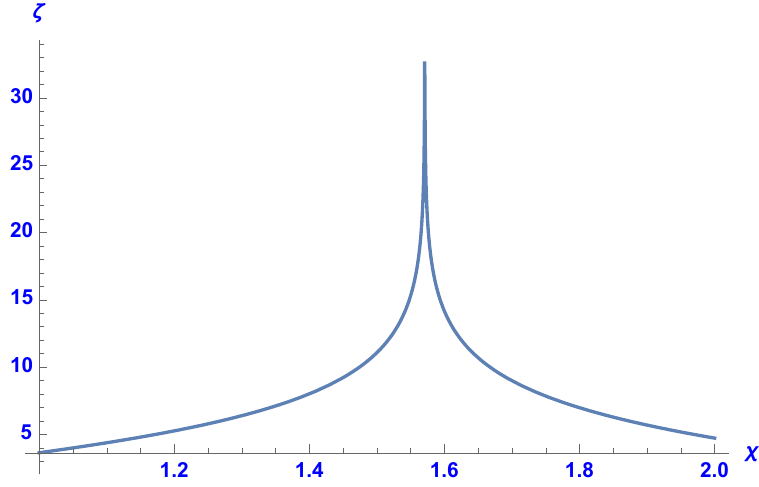}
\end{center}
  \caption{Single spike profile for ATD background where we set $ b=1 $ and $ a=2 $.} \label{atd}
\end{figure}

\paragraph{Single spikes.} Discussion for single spike solution follows analogous to that of \cite{Ishizeki:2007we}. For single spikes, one imposes the following boundary condition \cite{Park:2010vs}
\begin{align}
\label{e3.21}
\partial_{\sigma^0}\xi\Big|_{\chi = \chi_{max}}  =0
\end{align}
which yields the maximum angle associated with the string soliton to be
\begin{align}
\sin^2\chi_{max}=\frac{C a}{\omega b^2}.
\end{align}

Clearly, the infinite size limit of the spike corresponds to setting, $ C=\frac{\omega b^2}{a} $. This yields the minimum angle for the string soliton to be
\begin{align}
\label{e3.22}
\sin^2 \chi_{min}=\frac{b^2}{a^2}.
\end{align}

To proceed further, we first note down the energy of the string soliton
\begin{align}
\label{e3.23}
E = \frac{\sqrt{\lambda}a}{\pi}\int_{\chi_{min}}^{\pi/2}\frac{\cos^2 \chi_{min}\sin\chi}{\sin\chi_{min}\cos\chi \sqrt{\sin^2\chi - \sin^2\chi_{min}}}d\chi
\end{align}
where we set, $ \omega =1 $ and use \eqref{e3.22}.

The angular momentum, on the other hand, turns out to be
\begin{align}
J=\frac{\sqrt{\lambda}}{\pi}\int_{\chi_{min}}^{\pi/2}\frac{d\chi}{\chi'}\sin^2\chi \dot{\xi}=\frac{\sqrt{\lambda}b}{\pi}\int_{\chi_{min}}^{\pi/2}\frac{\sin\chi \cos\chi}{\sqrt{\sin^2\chi - \sin^2\chi_{min}}}d\chi=\frac{\sqrt{\lambda}b}{\pi}\cos \chi_{min}.
\end{align}

Finally, the angular difference turns out to be
\begin{align}
\label{e3.25}
\Delta \xi =2b \int_{\chi_{min}}^{\pi/2}\frac{\sin^2\chi -\sin^2 \chi_{min}}{\sin\chi_{min}\sin\chi \cos \chi \sqrt{\sin^2\chi -\sin^2 \chi_{min}}}d\chi.
\end{align}

A straightforward combination of the above integrals \eqref{e3.23}-\eqref{e3.25} reveals 
\begin{align}
\label{e3.27}
\tilde{E}=\frac{\sqrt{\lambda}}{\pi}\Delta \tilde{\xi}+\frac{\sqrt{\lambda}}{\pi}(\frac{\pi}{2}-\chi_{min})=\frac{\sqrt{\lambda}}{\pi}\Delta \tilde{\xi}+\frac{\sqrt{\lambda}}{\pi}\sin^{-1}\Big(\frac{\pi \tilde{J}}{\sqrt{\lambda}} \Big)
\end{align}
dispersion relation corresponding to spike \cite{Ishizeki:2007we}. Here, tildes define rescaled variables as
\begin{align}
\label{e3.28}
\tilde{E}=\frac{E}{a}~;~\tilde{\xi}=\frac{\xi}{2b}~;~\tilde{J}=\frac{J}{b}.
\end{align}
\subsection{Sfetsos-Thompson background}
\begin{figure}
\begin{center}
\includegraphics[scale=0.73]{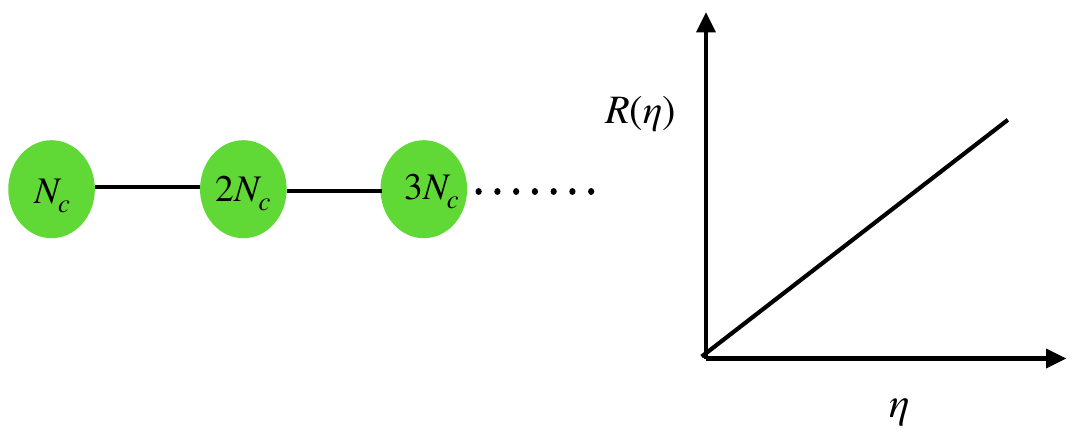}
\end{center}
  \caption{$ \mathcal{N}=2 $ quiver dual to ST background and the corresponding rank function.} \label{natd}
\end{figure}

The next example, that we focus upon, goes under the name of the ST solution which is obtained via NATD along $ SU(2) $ of the internal five sphere ($ S^5 $). The background is characterized by the potential function of the following form \cite{Lozano:2016kum}
\begin{align}
\label{st}
V_{ST}(\sigma, \eta)=\eta \log\sigma - \eta \frac{\sigma^2}{2}+\frac{\eta^3}{3}~;~R (\eta)=\eta
\end{align}
which represents $ \mathcal{N}=2 $ quiver with linearly increasing rank for the color group (see Fig.\ref{natd}).

Although the rank function $ R(\eta) $ vanishes at $ \eta =0 $ (see Fig.\ref{natd}), however, it does not truncate like in the ATD example. In other words, the dual quiver represents $ \mathcal{N}=2 $ SCFTs with infinite charge density. This also gets reflected in the corresponding Hanany-Witten (HW) set up which represents color D4 branes stretched in between NS5 branes with no flavor D6 branes to complete the geometry. In the next section, we cure this problem and discuss the completion of these quivers by adding flavor D6 branes into the story.

Using \eqref{st}, the corresponding metric functions could be found as
\begin{align}
&f_1(\sigma, \eta)=1~,~f_2(\sigma , \eta)=-\frac{4}{\sigma ^2-1}~,~f_3 (\sigma , \eta)=\frac{4 \eta  \left(\eta -\eta  \sigma ^2\right)}{4 \eta ^2+\left(\sigma ^2-1\right)^2}\\
&f_4(\sigma , \eta)=4 \sigma ^2~,~f_5(\sigma , \eta)=-\frac{8 \eta ^3}{4 \eta ^2+\left(\sigma ^2-1\right)^2}.
\end{align}

In what follows, we choose a string embedding of the following form
\begin{align}
\eta = \eta_c ~,~ \sigma = \sigma_c.
\end{align}

The corresponding Lagrangian density reads as
\begin{align}
\label{e3.33}
&\mathcal{L}_P = 4 \kappa^2  -\hat{f}_3 (a^2 -b^2) \chi'^2(\zeta)-\hat{f}_4 (a^2 -b^2) \beta'^2(\zeta) \nonumber\\
&- \hat{f}_3 \sin^2\chi ((\omega + a g'(\zeta))^2 -b^2 g'^2(\zeta)) 
\end{align}
where $ \hat{f}_i = f_i (\sigma_c , \eta_c) $ are constants and we ignore contribution due to $ B_2 $ field (as compared to the metric) which is an artefact of the semiclassical limit ($\lambda = \frac{L^4}{\alpha'^2} \gg 1$). In other words, when $L^2$ and $\alpha'$ factors are properly inserted \cite{Lozano:2016kum}, one can show that NS-NS contribution is always sub-dominant in a semiclassical ($\frac{1}{\sqrt{\lambda}}$) expansion \cite{Frolov:2002av}. This can be seen by rescaling $\eta = \frac{r}{\sqrt{\lambda}}$ \cite{Lozano:2016kum} and thereby taking a large $\lambda \gg 1$ limit. Under such approximations, one can show that $\eta = \eta_c$ is a solution to the full $\eta$-equation of motion in the domain where $\eta_c \gg1$.

The dynamics associated to $ \beta (\zeta) $ and $ g(\zeta) $ yields
\begin{align}
\label{e3.34}
\beta (\zeta)=\zeta ~;~g'(\zeta)=\frac{1}{(b^2 -a^2)}\Big( \omega a -\frac{C}{\hat{f}_3\sin^2 \chi} \Big).
\end{align}

Using \eqref{e3.8} and \eqref{e3.34}, one finds
\begin{align}
\label{e3.35}
\chi'^2(\zeta)=\frac{\omega^2 b^2}{(b^2 - a^2)^2 \sin^2 \chi}\Big( -\sin^4\chi +\frac{\hat{D}^2}{\hat{f}_3\omega^2 b^2}\sin^2\chi -\frac{C^2}{\hat{f}_3^2\omega^2 b^2}  \Big)
\end{align}
where we denote $\hat{D}^2=(4 \kappa^2 - (b+a)^2 \hat{f}_4)(b-a)^2+2 \omega b C$.

Comparing with \eqref{e3.9}, one can see the appearance of an additional factor $\hat{f}_3$ which characterizes the rank of the color $SU(P N_c)$ group. Based on this coefficient, we classify magnons corresponding to different color gauge group. An identical classification holds for spikes as well. 

One could further factorize \eqref{e3.35} following the spirit as in \eqref{e3.10}, where we identify
\begin{align}
& \sin^2 \chi_{max}+\sin^2\chi_{min}=\frac{\hat{D}^2}{\hat{f}_3\omega^2 b^2}\\
& \sin^2 \chi_{max}\sin^2\chi_{min}=\frac{C^2}{\hat{f}^2_3\omega^2 b^2 }.
\end{align}
\paragraph{Magnons.} We first search for the giant magnon solutions corresponding to ST background. Following a condition similar to \eqref{e3.13}, we find
\begin{align}
\label{e3.38}
\sin^2\chi_{max}=\frac{C}{\hat{f}_3 \omega a}=\frac{C}{\omega a}\Big( 1+\frac{1}{4 \eta^2_c} \Big)
\end{align}
where we set, $ \sigma_c=0 $ for simplicity.

Clearly, one recovers the ATD result \eqref{e3.15} in the limit $ \eta_c \rightarrow \infty $, which corresponds to color $ SU(P N_c) $ gauge group with large ($P N_c \gg 1$) rank. On the other hand, for color groups with finite rank, one never reaches the ``undefoemed'' giant magnon magnon condition $C= \omega a$ (see, \eqref{e3.15}), instead one finds a modified condition of the form
\begin{align}
\frac{C}{\omega a}=1-\frac{1}{4 \eta^2_c} <1.
\end{align}

On the other hand, using \eqref{e3.38}, the corresponding minimum angle turns out to be
\begin{align}
\sin^2\chi_{min}=\frac{a^2}{b^2}
\end{align}
irrespective of the rank of the associated gauge group.

A straightforward computation reveals the ``regularized'' magnon dispersion relation\footnote{The regularized dispersion relation is obtained by subtracting the divergent pice by adding an appropriate counter term $\sim \tanh ^{-1}(\sqrt{1-\epsilon^2})$, where $\epsilon \sim \cos \chi_{max} \sim 0$ in the strict giant magnon limit. The ideal giant magnon limit is defined first by taking a $\eta_c \rightarrow \infty$ limit and thereby setting, $\chi_{max}=\frac{\pi}{2}$ which gives, $ C=\omega a $.}
\begin{align}
\label{e3.41}
(E-J)_{reg}= \frac{\sqrt{\lambda}}{\pi} \sin \frac{p}{2}\Big( 1-\frac{ \Delta (p)}{4 P^2 N_c^2 }\Big)+\mathcal{O}(1/(P N_c)^4)
\end{align}
where $ p $ is the magnon momentum and $\eta_c \sim P N_c$ stands for the rank of the color group.

We interpret \eqref{e3.41} as the ``deformed'' giant magnon dispersion relation, which matches the ideal giant magnon spectrum \eqref{e3.20} in the large rank limit of the color group. Here, we identify LO contributions as an expansion of the following form
\begin{align}
\label{e3.42}
&\Delta (p)=  1+h(\epsilon, p)\nonumber\\
&h(\epsilon, p)=\frac{\epsilon^2}{4 \sin^2 \frac{p}{2}}+\mathcal{O}(\epsilon^4)~;~| \epsilon | \ll 1.
\end{align}

\begin{figure}
\begin{center}
\includegraphics[scale=0.73]{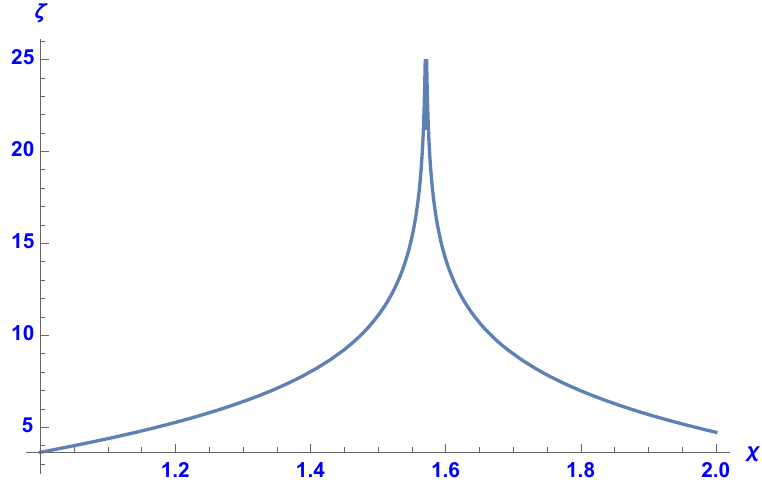}
\end{center}
  \caption{Single spike profile for NATD background where we set $ b=1 $, $ a=2 $ and $P N_c =500$.} \label{atd}
\end{figure}

\paragraph{Spikes.} The discussion for spikes follows quite analogous to that of the magnons. Using the boundary condition \eqref{e3.21}, one finds
\begin{align}
\sin^2\chi_{max}=\frac{C a}{\omega b^2 \hat{f}_3}.
\end{align}

Clearly, we deviate from the ``infinite size'' limit ($ C a = \omega b^2 $) of spikes and this turns out to be an artefact of the corrections appearing due to the finite rank of the color group
\begin{align}
\frac{C a}{\omega b^2}=1-\frac{1}{4 P^2 N^2_c} <1.
\end{align}
Clearly, for larger ranks, these corrections are suppressed and one approaches the ideal giant spike limit discussed in the previous section.

On a similar note, the minimum angle in the bulk turns out to be
\begin{align}
\sin^2 \chi_{min}=\frac{b^2}{a^2}.
\end{align}

The (rescaled) angular momentum \eqref{e3.28} of the string turns out to be
\begin{align}
\label{e3.46}
\tilde{J}=\frac{\sqrt{\lambda}\hat{f}_3}{\pi}\cos \chi_{min}.
\end{align}

Using \eqref{e3.46}, the modified spiky dispersion relation turns out to be
\begin{align}
\tilde{E}=\frac{\sqrt{\lambda}}{\pi}\Delta \tilde{\xi}+\frac{\sqrt{\lambda}}{\pi}\sin^{-1}\Big(\frac{\pi \tilde{J}}{\sqrt{\lambda}} \Big)+\frac{  \tilde{J}}{4P^2 N_c^2 \sqrt{1-\frac{\pi ^2 \tilde{J}^2}{\lambda }}}+\mathcal{O}(1/(P N_c)^4)
\end{align}
where all the sub-leading corrections are due to the finite rank of the color gauge group of the $ \mathcal{N}=2 $ quiver. These corrections act like a deformation to the ideal spiky condition of the string. Clearly, ideal spiky dispersion \eqref{e3.27} is recovered in the limit, $ P N_c \rightarrow \infty $. Therefore, the ideal magnon as well as spiky dispersions are deformed in the presence of finite rank corrections and the ideal dispersion relation is recovered in the limit of infinite rank of the color group.
\section{Adding flavors}
We now discuss the completion of $ \mathcal{N}=2 $ quivers by adding appropriate flavor branes into the picture. As an example, we discuss ``single kink'' solutions that can be thought of as the completion of ST (see Fig.\ref{kink}) by adding flavor D6 branes into the Hanany-Witten set up \cite{Lozano:2016kum}.

\begin{figure}
\begin{center}
\includegraphics[scale=0.73]{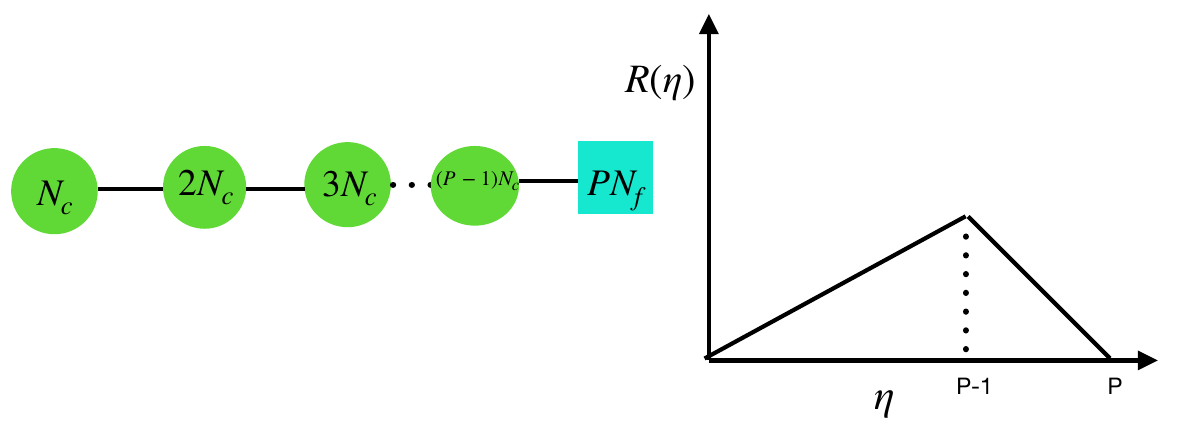}
\end{center}
  \caption{$ \mathcal{N}=2 $ quiver dual to single kink background and the corresponding rank function.} \label{kink}
\end{figure}

Potential function characterising the single kink profile could be expressed as \cite{Nunez:2018qcj}
\begin{align}
&\dot{V}_k(\eta , \sigma)=\frac{N_6}{2}\sum_{n=-\infty}^{\infty}(P+1)[\sqrt{\sigma^2 +(\eta +P -2n(1+P))^2}-\sqrt{(\eta -2 n (P+1)-P)^2+\sigma ^2}]\nonumber\\
&+P[\sqrt{(\eta -2 n (P+1)-P-1)^2+\sigma ^2}-\sqrt{(\eta -2 n (P+1)+P+1)^2+\sigma ^2}]
\end{align}
which represents $ N_6 $ flavor D6 branes sitting at, $ \eta =P $ of the internal manifold.

Considering $ -k\leq n \leq  k$ where $ k $ is a large number, an expansion near $ \sigma \sim 0 $ yields \cite{Roychowdhury:2021eas}
\begin{align}
\label{e4.3}
V_k (\sigma \sim 0, \eta)= \eta N_6 \log\sigma+\frac{\eta N_6 \sigma^2}{4}\Lambda_{k}(\eta , P)
-\frac{\eta N_6 \sigma^2}{4}\frac{(P+1)}{(P^2 -\eta^2)}
\end{align}
where we denote the above function as
\begin{align}
&\Lambda_{k}(\eta , P) =(P+1)\sum_{m=1}^{k}\Big(\frac{1}{(2m+(2m-1)P)^2-\eta^2}\nonumber\\
&-\frac{1}{(2m+(2m+1)P)^2-\eta^2}\Big)+\frac{P}{(2k+1)^2(1 +P)^2 -\eta^2}.
\end{align}

Using \eqref{e4.3}, the corresponding potential functions turn out to be
\begin{align}
\label{e4.4}
&f_1(\sigma \sim 0, \eta)= 4f^{-1}_2(\sigma \sim 0, \eta),\\
\label{e4.5}
&f_2(\sigma \sim 0, \eta)=\frac{\sqrt{2} }{\sqrt{F(P,\eta)}},\\
&f_3(\sigma \sim 0, \eta)=  \eta^2 f_2(\sigma \sim 0, \eta),\\
&f_4(\sigma \sim 0, \eta)=\sigma^2 f_2(\sigma \sim 0, \eta).
\label{e4.7}
\end{align}

We identify the above function in \eqref{e4.5} as
\begin{align}
F(P, \eta)=\frac{\eta  \left(P^2-\eta ^2\right)^3}{\sigma ^2 \left(\left(P^2-\eta ^2\right)^3 \left(\eta  \partial^2_{\eta}\Lambda_k (\eta , P )+2 \partial_{\eta}\Lambda_k (\eta , P )\right)-2 \eta  (P+1) \left(\eta ^2+3 P^2\right)\right)}
\end{align}

In what follows, we choose an embedding similar to that of the ST example which reveals a Lagrangian density as in \eqref{e3.33}. Here, $ \hat{f}_i (i=1, \cdots ,4)$s are precisely the functions \eqref{e4.4}-\eqref{e4.7} evaluated at $ \eta=\eta_c $ and $ \sigma = \sigma_c \ll 1$.
\paragraph{Magnons.} Following steps as mentioned before, the giant magnon limit yields 
\begin{align}
\label{e4.8}
\frac{C}{\omega a}=\eta_c^2 \hat{f}_2\Big|_{\eta_c \sim P}=\sqrt{\frac{2(P+1)}{P}}\frac{\sigma_c P^2}{(P-\eta_c)^{3/2}}\Big( 1+\frac{P-\eta_c}{2P} \Big).
\end{align}

Clearly, the giant magnon condition \eqref{e3.15} becomes ill-defined as the string approaches the singularity, $ \eta_c \sim P $ due to flavor branes. This is an artefact of ``fragmentation'' of closed strings \cite{Roychowdhury:2021eas} in the vicinity of flavor D6 branes, which states that closed strings get fragmented into smaller parts as they approach flavor D6 branes. In other words, a dispersion relation could be realised away ($ \eta_c \ll P $) from these flavor D6 branes, which for the present case reveals
\begin{align}
\frac{C}{\omega a}=\eta_c^2 \hat{f}_2\Big|_{\eta_c \ll P}=\frac{\eta^2_c \sigma_c}{2\sqrt{2}}\frac{\sqrt{K(P,k)}}{P^2 (1+P)^2 (1+2k)^2}
\end{align}
where we denote the above function in terms of Poly-Gamma functions $\psi^{(n)}(z)$
\begin{align}
&K(P,k)=96 \Big(16 k^4 (P+1)^5+32 k^3 (P+1)^5+24 k^2 (P+1)^5+8 k (P+1)^5\nonumber\\
&+5 P \left(P^2+P+1\right) (P+1)+1\Big)-(P+1) (2 k P+P)^4\Big(\psi ^{(3)}\left(\frac{P+2}{2 P+2}\right)-\psi ^{(3)}\left(\frac{3 P+2}{2 P+2}\right)\nonumber\\
&- \psi ^{(3)}\left(\frac{2 P k+2 k+P+2}{2 P+2}\right)+\psi ^{(3)}\left(\frac{2 P k+2 k+3 P+2}{2 P+2}\right) \Big).
\end{align}

A careful expansion for $ P \gg 1$, further reveals 
\begin{align}
\label{e4.12}
\frac{C}{\omega a}\Big|_{P \gg 1} = \frac{2\sqrt{3}\eta^2_c \sigma_c }{P^{3/2} }
\end{align}
where we also consider, $ P \gg k \gg 1 $.

A careful look reveals that the ideal giant magnon condition \eqref{e3.15} is not valid and instead one has $\frac{C}{\omega a}\ll 1$. Therefore, a deviation from ideal giant magnon dispersion relation is whatsoever expected. A careful analysis reveals a modified dispersion relation of the form 
\begin{align}
(E-J)_{reg}=\frac{\sqrt{\lambda}}{\pi}\eta_c^2 \hat{f}_2 \sin \frac{p}{2}\Big(1-\frac{1-\eta^2_c \hat{f}_2}{\eta^2_c \hat{f_2}}h(\epsilon,p)  \Big).
\end{align}

Here, $ h(\epsilon,p) $ is the function of magnon momentum that is defined in \eqref{e3.42}. Following our previous discussion \eqref{e4.8}, one could see that the giant magnon sates are ill-defined as we approach ($ \eta_c \sim P $) flavor D6 branes 
\begin{align}
E-J = \frac{\sqrt{\lambda}}{\pi} \frac{\sigma_c\sqrt{2P^3 (P+1)} }{(P-\eta_c)^{3/2}}\sin \frac{p}{2}(1+ h(\epsilon,p)).
\end{align}

On the other hand, using \eqref{e4.12}, away from flavor branes one finds
\begin{align}
\label{e4.15}
E-J = \frac{2\sqrt{3}\sqrt{\lambda}\eta^2_c \sigma_c}{\pi P^{3/2} }\sin \frac{p}{2}.
\end{align}

Therefore, the magnon spectrum \eqref{e4.15} clearly vanishes in the limit when $  \frac{\eta^2_c \sigma_c}{P^{3/2}}\ll\frac{1}{\sqrt{\lambda}} \ll 1$. This clearly shows that magnons are ceased to exist even away from the flavor branes. 
\paragraph{Spikes.} The analysis for spikes follows in a way similar to that of the ST example. The (re-scaled) angular momentum is given by
\begin{align}
\label{e4.13}
\tilde{J}=\frac{\sqrt{\lambda}}{\pi}\eta^2_c\hat{f}_2\cos \chi_{min}.
\end{align}

Using \eqref{e4.13}, one finds a deformed dispersion relation of the form
\begin{align}
\tilde{E}=\frac{\sqrt{\lambda}}{\pi}\Delta \tilde{\xi}+\frac{\sqrt{\lambda}}{\pi}\sin^{-1}\Big(\frac{\pi \tilde{J}}{\sqrt{\lambda}\eta^2_c\hat{f}_2} \Big).
\end{align}

Furthermore, an expansion near flavor branes ($ \eta_c \sim P \gg 1$) reveals 
\begin{align}
\label{e4.16}
\tilde{E}=\frac{\sqrt{\lambda}}{\pi}\Delta \tilde{\xi}+\frac{\tilde{ J} (P-\eta_c) ^{3/2}}{\sqrt{\frac{2}{P}+2} P^2 \sigma_c }
\end{align}
a non-spiky spectrum \cite{Ishizeki:2007we}. 

Close to flavor branes, one can ignore the second term on the r.h.s. of \eqref{e4.16}, which reveals
\begin{align}
\tilde{E}=\frac{\sqrt{\lambda}}{\pi}\Delta \tilde{\xi}.
\end{align}

On the other hand, using \eqref{e4.12}, away from the flavor branes one finds 
\begin{align}
\label{e4.20}
\tilde{E}=\frac{\sqrt{\lambda}}{\pi}\Big(\Delta \tilde{\xi}+\frac{\pi \tilde{J} P^{3/2}}{2 \sqrt{3}\sqrt{\lambda} \eta_c ^2 \sigma_c }\Big)
\end{align}
a ``non-spiky'' spectrum as well where we retain ourselves up to LO corrections only. 

Using \eqref{e4.12} and \eqref{e4.13}, one could further simplify \eqref{e4.20} to yield
\begin{align}
\tilde{E}=\frac{\sqrt{\lambda}}{\pi}\Big(\Delta \tilde{\xi}+\cos \chi_{min}\Big).
\end{align}

Clearly, we have a small correction factor $ |\cos \chi_{min}|\leq 1 $ away from the flavor brane. This ensures the overall absence of spikes in the presence of flavor degrees of freedom.

Therefore, to summarise, unlike the ST (or the ATD), one looses the spiky dispersion relation in the presence of flavor branes. Our analysis confirms that $ \mathcal{N}=2 $ quivers with added flavor possesses excitations that are different from that of rest of the class where the flavor is missing.

\begin{figure}
\begin{center}
\includegraphics[scale=0.7]{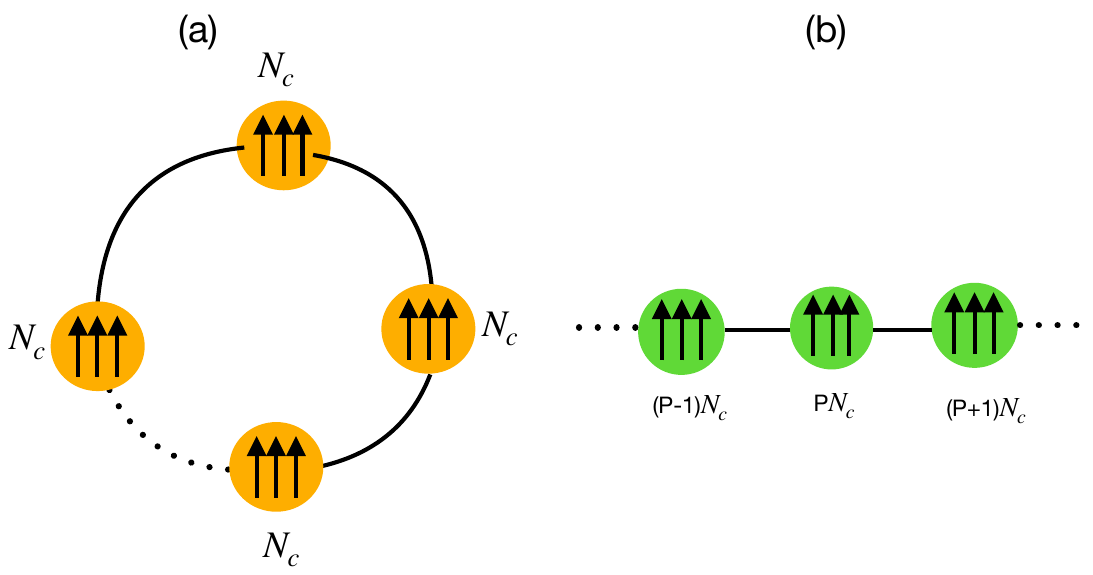}
\end{center}
  \caption{(a) $ \mathcal{N}=2 $ quiver dual to ATD background that could be recast as a collection of magnon excitations of a spin chain in the semiclassical limit. This forms a circular array of spin chains. Here, in the figure each arrow corresponds to some local excitation of the spin chain. (b) $ \mathcal{N}=2 $ quiver dual to NATD/ST background that could be modeled as a linear array of spinchains where each color node is associated with magnon like excitations.} \label{kink}
\end{figure}
\section{Conclusions and Outlook}
We now summarise the key findings of our paper. In the present paper, we explore possibilities of magnon and single spike solutions for $ \mathcal{N}=2 $ Gaiotto-Maldacena (GM) backgrounds in type IIA supergravity. As a trivial example, we notice that these solutions indeed exist for Abelian T-dual (ATD) geometries. On the other hand, for Sfetsos-Thompson (ST) background, we have a class of ``deformed'' giant magnon dispersion relations that are characterized by the rank function of the associated color gauge group. These results boil down into the standard giant magnon dispersion \cite{Hofman:2006xt}, in the limit when the rank of the $SU(P N_c)$ color nodes in $ \mathcal{N}=2 $ quiver becomes infinitely large. Similar remarks hold for spikes as well. 

When flavor D6 branes are added, giant magnon states are well defined only in the regime where the string is away from flavor D6 branes. We identify this as an artefact of ``fragmentation'' of heavy operators \cite{Roychowdhury:2021eas} in $ \mathcal{N}=2 $ SCFTs. As the string approaches flavor branes near $ \eta_c \sim P $, it gets fragmented into smaller strings. This is reflected as the fragmentation of long single trace operators in $ \mathcal{N}=2 $ SCFTs, which could be schematically expressed as
\begin{align}
\mathcal{O}^{f_m f_{m+n}}\sim \mathcal{O}^{f_m }\mathcal{O}^{ f_{m+n}}
\end{align}
where $ f_m $ are the flavor indices. Clearly, the operators $  \mathcal{O}^{f_m } $ are individually dual to fragmented string states. Our analysis reveals that magnon excitations cannot propagate through the quiver when flavor nodes are present. Flavor nodes impose ``defect'' along the spin chain, which ceases the magnon modes/ spin chain excitations to propagate further. This is related to the fact that a long single trace operator $ \mathcal{O} $ is ill-defined and gets broken apart due to flavor excitations. 

Therefore, the take home message is the following- neither magnons nor spikes are well defined entities in the presence of flavor degrees of freedom. They can only be defined for an infinitely long $ \mathcal{N}=2 $ linear quiver. This picture is quite intriguing if we compare with previous analysis of \cite{Nunez:2018qcj}, which shows that $ \mathcal{N}=2 $ backgrounds are (Liouvillian) integrable only in the absence of flavor branes. Combining the above pictures together, one could think of the ATD model analogous to ``integrable'' spin chain which possesses magnons as fundamental excitations. In the dual stringy picture, these excitations propagate along the string world-sheet where the string wraps the holographic $ \eta $- direction. On the other hand, the ST and/or NATD background are dual to an infinite ``array'' of integrable spin chains, where each spin chain in the array has its own excitations that could be identified with ``modified'' magnon dispersion.

The above formalism could be tested for the other classes of holographic dualities that are constructed in the recent years. One of these models correspond to $\mathcal{N}=1$ SCFTs in five dimensions and their holographic type IIB duals with an $AdS_6$ factor \cite{DHoker:2016ujz}-\cite{DHoker:2017zwj}. An electrostatic version of balanced $\mathcal{N}=1$ quivers were recently constructed by authors in \cite{Legramandi:2021uds} which allows for a potential function $V(\sigma , \eta)$ formulation analogous to the NATD model discussed in this paper. These quivers therefore allow for one more application of the computational techniques presented in the present manuscript. The second example corresponds to $\mathcal{N}=(0,4)$ SCFTs in two dimensions that are dual to massive type IIA solutions with an $AdS_3$ factor \cite{Lozano:2019emq}-\cite{Lozano:2019zvg}. The bulk solution corresponds to a piece-wise linear metric profile where the methods of the present paper could be applied for each interval of the metric function.
\paragraph {Acknowledgements.}
 The author is indebted to the authorities of IIT Roorkee for their unconditional support towards researches in basic sciences. Its a pleasure to thank Carlos Nunez for useful conversations. The author would also like to acknowledge The Royal Society, UK for financial assistance. The author would also like to acknowledge the Mathematical Research Impact Centric Support (MATRICS) grant (MTR/2023/000005) received from SERB, India.

\end{document}